\documentclass[twocolumn,twoside]{IEEEtran} \def\twocol{}
\usepackage{graphicx,amssymb,amsmath}
\usepackage{subfigure}
\usepackage[latin1]{inputenc}

\def\B{{\bf B}}
\def\b{{\bf b}}
\def\A{{\bf A}}
\def\E{{\rm E}}
\def\X{{\bf X}}
\def\S{{\bf S}}
\def\Y{{\bf Y}}
\def\U{{\bf U}}
\def\u{{\bf u}}
\def\W{{\bf W}}
\def\w{{\bf w}}
\def\I{{\bf I}}
\def\T{{\rm T}}
\def\H{{\cal H}}
\DeclareMathOperator{\e}{e}
\DeclareMathOperator{\erf}{Erf}
\DeclareMathOperator{\erfc}{Erfc}
\DeclareMathOperator{\var}{var}

\newtheorem{lemma}{Lemma}
\newtheorem{coroll}{Corollary}
\newtheorem{example}{Example}

\def\qed{\hbox{\hskip1pt\vrule width4pt height6pt depth1.5pt}}

\title{Mixing and non-mixing local minima of the entropy contrast for blind source separation}
\author{Frédéric Vrins\IEEEmembership{,~Student Member,~IEEE}, Dinh-Tuan Pham\IEEEmembership{,~Member,~IEEE} and Michel Verleysen\IEEEmembership{,~Senior~Member,~IEEE}}
\begin{document}
\maketitle

\begin{abstract}

In this paper, both non-mixing and mixing local minima of the
entropy are analyzed from the viewpoint of blind source separation
(BSS); they correspond respectively to acceptable and spurious
solutions of the BSS problem. The contribution of this work is
twofold. First, a Taylor development is used to show that the
\textit{exact} output entropy cost function has a non-mixing
minimum when this output is proportional to \textit{any} of the
non-Gaussian sources, and not only when the output is proportional
to the lowest entropic source. Second, in order to prove that
mixing entropy minima exist when the source densities are strongly
multimodal, an entropy approximator is proposed. The latter has
the major advantage that an error bound can be provided. Even if
this approximator (and the associated bound) is used here in the
BSS context, it can be applied for estimating the entropy of any
random variable with multimodal density.
%
\end{abstract}
\begin{keywords}
Blind source separation. Independent component analysis. Entropy
estimation. Multimodal densities. Mixture distribution.
\end{keywords}

\begin{center} \bfseries EDICS Category: \end{center}

\section{Introduction}

Blind source separation (BSS) aims at recovering a vector of
independent sources $\S = [S_1, \cdots, S_K]^\T$ from observed
mixtures $\X = [X_1, \cdots, X_M]^\T$. In this paper, we assume
that $K=M$ and $\X=\A\S$, where $\A$ is the $K$-by-$K$ mixing
matrix. The sources can be recovered by finding an unmixing matrix
$\B$ such that $\W=\B\A$ is non-mixing (i.e. with one non-zero
entry per row and per column). Such matrices $\B$ can be found by
minimizing an ad-hoc cost function (see \cite{cit_Como94}, the
books \cite{cit_ICABook,cit_BSS,cit_CichokiBook} and references
therein).

In practice, the minimum of these criteria is reached by adaptive
methods such as gradient descents. Therefore, one has to pay
attention to the solutions corresponding to these minima. In most
of cases, the global minimum is a solution of the BSS problem. By
contrast, the possible local minima can either correspond to a
desired solution (referred as \emph{non-mixing} minima) or
spurious solution (referred as \emph{mixing} minima) of the
problem. For example, the optimization algorithm could be trapped
in minima that do not correspond to an acceptable solution of the
BSS problem. Therefore, it is of interest to study the possible
existence of both non-mixing and mixing local minima.

The paper deals with this issue by extending existing results of
related work. The introduction first presents the two main
approaches for source separation and details the state-of-the-art
related to the local minima of BSS criteria. Then, the objectives
and the organization of the paper is presented.

\subsection{Symmetric and deflation approaches}

To determine matrix $\B$, two approaches can be investigated. The
first one (called \textit{symmetric}) aims at extracting all
sources simultaneously. The second approach (called
\textit{deflation}) extracts the sources one by one.

\begin{itemize}
\item The common symmetric approach consists in minimizing the
Kullback-Leibler divergence between the joint density and the
product of the marginal densities of the recovered sources (i.e.
their mutual information), which are the components $Y_1, \dots,
Y_K$ of $\Y = \B\X$. This leads to the minimization of (see
\cite{cit_MI,cit_GrayIT,cit_PhamIEEE_IT})
\begin{equation}
\label{Cont} C(\B) = \sum_{i=1}^K H(Y_k) - \log|\det\B|\enspace,
\end{equation}
where $H(Y)$ denotes Shannon's differential entropy
$Y$~\cite{cit_MI,cit_GrayIT}:
\begin{equation}
H(Y)=-\int p_Y(y)\log(p_Y(y))dy\enspace. \label{Def:Shannon}
\end{equation}
In eq. (\ref{Def:Shannon}), $p_Y$ denotes the probability density
function (pdf) of $Y$. A variant of this approach applies the
unmixing matrix $\B$ to a whitened version of the observations. In
this case, since the sources are uncorrelated and can be assumed
to have the same variance, one can constrain $\B$ to be orthogonal
\cite{cit_ICABook}. The term $\log\det\B$ in
criterion~(\ref{Cont}) disappears and $C(\B)$ is to be minimized
over the group of orthogonal matrices.

\item The deflation approach~\cite{cit_Deflation} extracts the
$k$-th source by computing the $k$-th row $\b_k$ of $\B$ by
minimizing a non Gaussianity index of $\b_k\X$ subject to the
constraint that $\b_k\X$ is uncorrelated to ${\bf b}_i \X$ for $i
< k$. By taking this index to be the negentropy \cite{cit_FastICA}
and assuming (without loss of generality) that the sources have
the same variance, the cost function can be written as $H({\bf
w}_k\S) - \log\|{\bf w}_k\|$ plus a constant, where ${\bf w}_k =
{\bf b}_k\A$ and $\|{\bf w}_k\|$ denotes the Euclidean norm
$\sqrt{{\bf w}_k{\bf w}_k^T}$ \cite{cit_Cruces-IEEE-NN,
cit_PhamIEESP_ICA}. Since this function is unchanged when ${\bf
w}_k$ is multiplied by a scalar, this leads to minimizing $H({\bf
w}_k\S)$ under the ${\bf w}_i {\bf w}_k^\T = \delta_{i,k}$
constraint for $1\leq i,k\leq K$, where $\delta_{i,k}$ is the
Kronecker delta \cite{cit_Gray}.

\end{itemize}

\subsection{Related works}

Although both symmetric and deflation procedures could be analyzed
in this contribution with the same tools, we focus on the entropy
$H(Y_k)$, used in the deflation approach.

Several results exist regarding the entropy minima of $Y = \w\S$
(the subscript ``$k$'' has been omitted in the following, since
one signal is extracted at a time in the deflation approach). The
first kind of results discusses the existence of non-mixing local
minima of $H(Y)$ that correspond to the extraction of a single
source. The second kind of results discusses the existence of
mixing minima that correspond to spurious solutions of the BSS
problem: $Y$ is still a mixture of sources despite the fact that
$H(Y)$ is a local minimum. These results are summarized below.

\begin{itemize}

\item \textit{Non-mixing entropy local minima}\\
It has been shown that the global minimum of $H(Y)$ with $Y =
\w\S$ is reached when the output $Y$ is proportional to the source
with the lowest entropy \cite{cit_Cruces-IEEE-NN}. It is proven in
\cite{cit_FastICA} that when a fixed-variance output is
proportional to one of the sources, then, under some technical
conditions, the cumulant-based approximation of entropy $H_J(Y)$
used in FastICA \cite{cit_FastICA} reaches a non-mixing local
minimum. Finally, based on the entropy power inequality
\cite{cit_Dembo}, it is also proven in \cite{cit_vrins_SP_04}
that, in the two-dimensional case, Shannon's entropy has a local
minimum when the output is proportional to a non-Gaussian source.

\item\textit{Mixing entropy local minima}\\
As for the mutual information, simulations results
in~\cite{cit_vrins_IEEESPL_05a} suggest that mixing local entropy
minima exist in specific cases (i.e. when the source pdfs are
strongly multimodal, which sometimes occur in practice, for
sinusoid waveforms among other). These results, based on density
estimation using the Parzen kernel method, are confirmed by other
simulations using directly entropy estimation, such as Vasicek's
one in \cite{cit_learned-miller_04} or based on the approximator
analyzed in this paper in \cite{cit_vrins_Eusipco05_a}. Rigorously
speaking, the above results do not constitute an absolute proof
since error bounds are not available for the approximation
procedure. By contrast, a theoretical proof is given in
\cite{cit_vrins_IEEESPL_05b}, but for a specific example only (two
bimodal sources sharing the same symmetric pdf). The existence of
mixing local entropy minima has also been shown
in~\cite{cit_PhamVrins_ISPPA05} (without detailed proof) in the
case of two non symmetric sources with strongly multimodal pdfs.

\end{itemize}

\subsection{Objectives and organization of the paper}
In this paper, additional results regarding mixing and non-mixing
entropy minima are presented. Two main results will be proven.

Firstly, it will be shown in the next section that the exact
entropy of an output $H(Y)$ with a fixed variance has local
non-mixing minima: the entropy $H(Y)$ has a local minimum when $Y$
is proportional to one of the non-Gaussian sources. This is an
extension of the results presented in \cite{cit_vrins_IEEESPL_05b}
to the case of $K > 2$ sources. If the output is proportional to
the Gaussian source (if it exists), the entropy has a global
maximum. Numerical simulations illustrate these results in the
$K=2$ case, for the ease of illustration.

Secondly, in Section III, an entropy approximator is presented,
for which an error bound can be derived. It is suitable for
variables having multimodal densities with modes having a low
overlap, in the sense that its error bound converges to zero when
the mode overlap becomes negligible. This approximator was
mentioned in \cite{cit_vrins_Eusipco05_a} and error bounds have
been provided in~\cite{cit_PhamVrins_ISPPA05} without proof. In
the BSS context, when the sources have such densities, the use of
this approximator makes it possible to show that the marginal
entropy has local mixing minima. This approach can be applied to a
wider class of source densities than the score function-based
method derived in \cite{cit_vrins_IEEESPL_05b}. The results
presented in this paper further extend those
in~\cite{cit_PhamVrins_ISPPA05} as they are not restricted to the
case of $K = 2$ sources. Finally, we provide a detailed proof of
the bound formula for the entropy approximator.

It must be stressed that the aforementioned entropy approximator
can be used for other applications that require entropy estimation
of multimodal densities.

\section{Local non-mixing minima of output entropy}
In this section, we shall prove that $H(\w\S)$, under the $\|\w\|
= 1$ constraint, reaches a local minimum at $\w=\I_j$, the $j$-th
row of the $K \times K$ identity matrix, if $S_j$ is non-Gaussian,
or a global maximum otherwise. Note that, as it is well known, the
global minimum is reached at $\I_k$ where $k = \arg\min_k H(S_k)$.

\subsection{Theoretic development}

The starting point is an expansion of the entropy of a random
variable $Y$ slightly contaminated with another variable $\delta
Y$ up to second order in $\delta Y$, which has been established
in~\cite{cit_PhamEntrop}:
\begin{eqnarray}
H(Y + \delta Y) &{}\approx{}& H(Y) + \E[\psi_{Y}(Y)\delta Y]
+\nonumber\\
&& {1\over2} \{\E[\var(\delta Y | Y) \psi_Y'(Y)] - [\E(\delta Y |
Y)]^{\prime\,2}\}\enspace. \label{e:EntropDev}
\end{eqnarray}
In this equation, $\psi_Y$ is the score function of $Y$, defined
as $-(\log p_Y)'$\footnote{In this paper, we use the score
function definition presented in \cite{cit_PhamIEEE_IT}. However,
several authors define this function with the opposite sign. The
reader should have this difference in mind.}, $p_Y$ is the pdf of
$Y$, $'$ denotes the derivative, and $\E(\cdot|Y)$ and
$\var(\cdot|Y)$ denote the conditional expectation and conditional
variance given $Y$, respectively.

Assume that $\w$ is close from $\I_j$ so that its $i$-th component
$w_i$ is close to 0 for $i \ne j$. Under the $\|\w\|=1$
constraint, $w_j = \sqrt{1-\sum_{i\neq j} w_i^2}$ and since
$\sqrt{1-x} = 1-\frac{1}{2}x + o(x)$, one can write
\[
w_j=1-\frac{1}{2}\sum_{i\neq j} w_i^2 + o\Big(\sum_{i\neq j} w_i^2\Big).
\]
Thus, $\w\S = S_j + \delta S_j$ with
\[
\delta S_j = \sum_{i\neq j} w_i S_i -
\frac{1}{2} \Big(\sum_{i\neq j} w_i^2\Big) S_j +
o\Big(\sum_{i\neq j} w_i^2\Big).
\]
Therefore, applying (\ref{e:EntropDev}) and dropping higher order
terms, one gets that $H(\w\S)$ equals
\[
\displaylines{
H(S_j) + \Big(\sum_{i\neq j} w_i\Big) \E[\psi_{S_j}(S_j) S_i] -
{1\over2}\Big(\sum_{i\neq j} w_i^2\Big) \E[\psi_{S_j}(S_j) S_j]
\hfill\cr\hfill
{}+ \frac{1}{2}
\Big\{\E\Big[{\rm var}\Big(\sum_{i\neq j} w_i^2 S_i \Big| S_j\Big)
\psi_{S_j}'(S_j)\Big] -
\Big[\sum_{i\neq j} w_i \E(S_i | S_j)\Big]^{\prime\,2}\Big\}
\cr\hfill
{}+ o\Big(\sum_{i\neq j} w_i^2\Big).
}
\]

Since the sources are mutually independent, any non-linear mapping
of them is uncorrelated so that $\E[\psi_{S_j}(S_j) S_{i}] = 0$,
for $i \neq j$. Furthermore $\E(S_{i}|S_j) = \E(S_i) = 0$ for
$i\neq j$,  $\E[\psi_{S_j}(S_j) S_{j}]=1$ (by integration by
parts), and ${\rm var}(\sum_{i \ne j} w_i S_i|S_j) = {\rm
var}(\sum_{i\ne j} w_i S_i) = (\sum_{i \ne j} w_i^2) \sigma_S^2$
where $\sigma_S^2$ denotes the common variance of the sources.
Therefore
\begin{eqnarray}
\nonumber
H(\w\S) &{}={}& H(S_j) +
\frac{1}{2} \Big(\sum_{i\neq j} w_i^2\Big)
\{\sigma_S^2\E[\psi_{S_j}'(S_j)] - 1\}
\\
&&\hspace{4cm}+ o\Big(\sum_{i \ne j} w_i^2\Big) .
\label{e:EntropDevFinal}
\end{eqnarray}
Note that again by integration by parts, $\E[\psi_{S_j}'(S_j)]$
can be rewritten as $\E[\psi_{S_j}^2(S_j)]$, which is precisely
Fisher's information~\cite{cit_MI}. In addition, by Schwarz's
inequality \cite{cit_MI}, one has
\[
|\E\{[S_j - \E(S_j)]\psi_{S_j}(S_j)\}| \le \sqrt{\sigma_S^2
\E[\psi_{S_j}^2(S_j)]}\enspace,
\]
with equality if and only if $\psi_{S_j}$ is a linear function.
But since as mentioned above $\E[\psi_{S_j}(S_j)] = 0$ and
$\E[S_j\psi_{S_j}(S_j)] = 1$, the left hand side of the above
inequality equals 1. Thus $\sigma_S^2 \E[\psi_{S_j}^2(S_j)] > 1$
unless $\psi_{S_j}$ is linear (which means that $S_j$ is Gaussian)
in which case $\sigma_S^2 \E[\psi_{S_j}^2(S_j)] = 1$. One
concludes from~(\ref{e:EntropDevFinal}) that $H(\w\S) > H(S_j)$
for all $\w$ sufficiently close to $\I_j$ if $S_j$ is
non-Gaussian. Thus $H(\w\S)$ reaches local non-mixing minima at
$\w = \pm\I_j$ (since $H(-\w\S) = H(\w\S)$), as long as $S_j$ is
non-Gaussian. If $S_j$ is Gaussian then $H(S_j)$ is a global
maximum since Gaussian random variables have the highest entropy
for a given variance. Equality~(\ref{e:EntropDevFinal}) is of no
use in this case, since the second term in this equality vanishes.

\subsection{Numerical simulations}
In this subsection, three simple examples are analyzed in the
$K=2$ case. In this case, the unit-norm vector $\w$ can be
rewritten as $[\sin\theta,~\cos\theta]$ and $H(\w\S)$ is
considered as a function of $\theta$. The entropy is computed
through eq.~(\ref{Def:Shannon}), in which the pdf were estimated
from a finite sample set (1000 samples), using Parzen density
estimation \cite{Silverman1986,Scott92} with Gaussian Kernels of
standard deviation $\sigma_K=0.5\hat\sigma_X*S^{-1/5}$ ($S$
denotes the number of samples and $\hat\sigma_X$ is the empirical
standard deviation, enforced to be equal to one here) and
Riemannian summation instead of exact integration.

\begin{figure}
  \centering
\includegraphics[width=8.5truecm]{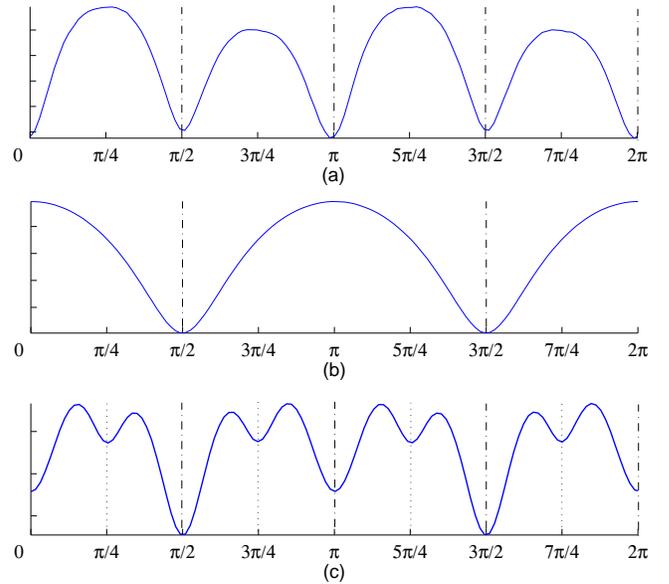}
 \caption{Evolution of of $H(\w\S)$ vs
$\theta$: (a) Example 1: two Uniform sources (b) Example 2:
Uniform ($S_1$) and Gaussian ($S_2$) sources; (c) Example 3: two
bimodal sources. The non-mixing minima are indicated by
dash-dotted vertical lines, the mixing ones by dotted
lines.}\label{Fig:Examples}
\end{figure}

\begin{example}
Assume that $S_1$ and $S_2$ have uniform densities. According to
the above results, local minima exist for
$\theta\in\{p\pi/2|p\in\mathbb{Z}\}$. In this example, no mixing
minimum can be observed (Fig. \ref{Fig:Examples}(a)).
\end{example}

\begin{example}
Suppose now that $S_1$ and $S_2$ have uniform and Gaussian
distributions respectively. Local minima are found for
$\theta\in\{(2p+1)\pi/2\}$, $p\in\mathbb{Z}$, and local maxima for
$\theta\in\{p\pi\}$ (Fig. \ref{Fig:Examples}(b)). Again, no
spurious minimum can be observed in this example.
\end{example}

\begin{example}
Consider two source symmetric pdfs $p_{s_1}$ and $p_{s_2}$ that
are constituted by i) two non-overlapping uniform modes and ii)
two Gaussian modes with negligible overlap, respectively. One can
observe that non-mixing solutions occur for $\theta\in\{p\pi/2\}$
(Fig. \ref{Fig:Examples}(c)).
\end{example}

In addition to an illustration of the above theoretical result,
the last example shows the existence os spurious (mixing) local
minima for $\theta\notin\{p\pi/2\}$. However, the figure does not
constitute a proof of the existence of local minima of $H(\w\S)$;
the minima visible on the figure could indeed be a consequence of
the entropy estimator (more precisely, of the pdf estimation). In
the next section, we derive an entropy estimator and an associated
error bound. This approximator is efficient for estimating the
entropy of variables having multimodal densities, in the sense
that the error bound tends to zero when the mode overlaps
decrease. Next, thanks to this approximator, it will be
theoretically proven that mixing local minima exist for strongly
multimodal source densities.

\section{Entropy approximator}
\label{s:EntropEst} In this section, we introduce the entropy
approximator first derived in \cite{cit_vrins_Eusipco05_a}. The
detailed proofs of the upper and lower bounds of the entropy based
on this approximator, already mentioned in
\cite{cit_PhamVrins_ISPPA05} without proof, are given.
Illustrative examples are further provided. The entropy bounds
will be used in the next section to prove that for a specific
class of source distributions, the entropy function $H(\w\S)$ can
have a local minimum that does not correspond to a row of the
identity matrix. The presented approach yields more general
results than those in \cite{cit_vrins_IEEESPL_05b}, since it is no
longer constrained that the sources share a common symmetric pdf.

This approach relies on an entropy approximation of a multimodal
pdf of the form
\begin{equation}
\label{e:density}
p(y) = \sum_{n=1}^N \pi_n K_n(y) ,
\end{equation}
where $N>1$, $\pi_1, \dots, \pi_N$ are (strictly positive)
probabilities summing to 1 and $K_1, \dots, K_N$ are unimodal
pdfs. We focus on the case where the supports of the $K_n$ can be
nearly covered by disjoint subsets $\Omega_n$ ($n = 1, \dots, N$)
so that $p$ is strongly multimodal (with $N$ modes). In this case
a good approximation to the entropy of a random variable of
density $p$ can be obtained; this entropy will be {\em abusively}
denoted by $H(p)$ instead of $H(Y)$ where $Y$ is a random variable
with pdf $p$. Such approximation will be first derived informally
(for ease of comprehension) and then a formal development giving
the error bounds of the approximator is provided.

\subsection{Informal derivation of entropy approximator}
If the random variable has a pdf of the form (\ref{e:density}), then
its entropy equals
\begin{equation}
H(p) = -\int_{-\infty}^\infty \sum_{n=1}^N\pi_n K_n(y)\log
\Big[\sum_{n=1}^N \pi_n K_n(y)\Big]dy\enspace. \label{e:EntropPy}
\end{equation}

Suppose that there exists disjoint sets $\Omega_1, \dots,
\Omega_N$ that \textit{nearly} cover the supports of the $K_n$
densities; even if the $K_n$ have a finite support, the $\Omega_n$
may differ from the true support of the $K_i$ since these supports
may be not disjoint. Then, assuming that $\pi_n K_n(y) \geq 0$ is
small or zero for all $y\notin\Omega_n$ and noting that $0\log
0=0$ by convention (more rigorously: $\lim_{x\rightarrow 0^+}
x\log x=0$), one gets
\begin{eqnarray*}
H(p) &{}\approx{}& -\sum_{m=1}^{N} \int_{\Omega_m} \sum_{n=1}^N
\pi_n K_n(y)\log \Big[\sum_{n=1}^N \pi_n K_n(y)\Big]dy
\\
&{}\approx{}& -\sum_{m=1}^N \pi_m \int_{\Omega_m} K_m(y)
\log[\pi_m K_m(y)] dy\enspace.
\end{eqnarray*}
If we note $\pi = [\pi_1, \cdots, \pi_n]$ and
$h(\pi) \triangleq -\sum_{n=1}^N \pi_n \log \pi_n$ the entropy
of a discrete random variable taking $N$ distinct values with
probabilities $\pi_1, \dots, \pi_N$, then $H(p) \approx \H(p)$ where
\begin{equation}
\H(p) \triangleq \sum_{n=1}^N \pi_n H(K_n) + h(\pi) .
\label{Eq:ApproxSumEntrop}
\end{equation}

\subsection{Upper and lower bounds of the entropy of a multimodal
distribution}
\label{Sec:EntropBounds}

The entropy approximator $\H(p)$ in previous subsection is
actually an upper bound for the entropy. This claim is proved in
the following; in addition, a lower bound of the entropy will be
further provided. These bounds permit to analyze how accurate is
the approximation $H(p) \approx \H(p)$; they are explicitly
computed when all $K_n$ are Gaussian kernels.

\subsubsection{General results}
\label{GeneralCase}
The following Lemma provides upper and lower bounds for the entropy.

\begin{lemma}
\label{lemma1}

Let $p$ be given by~(\ref{e:density}), then

\begin{equation}
H(p) \le \H(p)\label{e:UpperBound}
\end{equation}
where $\H(p)$ is given by~(\ref{Eq:ApproxSumEntrop}).

In addition, assume that $\sup K_n = \sup_{y \in \mathbb{R}}
K_n(y) < \infty$ ($1 \le n \le N$) and let $\Omega_1, \dots,
\Omega_N$ be disjoint subsets which approximately cover the
supports of $K_1, \dots, K_N$, in the sense that
\begin{eqnarray}
\left \{
\begin{array}{lll}
\epsilon_n &{}\triangleq{}& \int_{\mathbb{R}\setminus\Omega_n} K_n(y)
dy\enspace,
\nonumber\\
&&\nonumber\\
\epsilon_n' &{}\triangleq{}& \int_{\mathbb{R}\setminus\Omega_n} K_n(y)
\log\frac{\sup K_n}{K_n(y)} dy
\end{array}\right.
\end{eqnarray}
are small. Then, we have
\begin{eqnarray}
H(p) &{}\ge{}& \H(p) - \sum_{n=1}^N \pi_n \epsilon_n'
\nonumber\\
&& - \sum_{n=1}^N \pi_n \Big[ \log\Big(\frac{\max_{1 \le m \le N}
\sup K_m}{\pi_n \sup K_n}\Big) + 1\Big] \epsilon_n .
\label{e:LowerBound}
\end{eqnarray}

\end{lemma}

The proof of this Lemma is given in Appendix I.

Let us consider now the case where the densities $K_n$
in~(\ref{e:density}) all have the same form:
\begin{equation}
\label{e:kernel} K_n(y) = (1/\sigma_n)K[(y-\mu_n)/\sigma_n]
\end{equation}
where $K$ is a bounded density of finite entropy. Hence
$H(K_n)=H(K)+\log\sigma_n$ and the upper
bound~(\ref{Eq:ApproxSumEntrop}) becomes
\begin{equation}
H(p) \le \H(p) = H(K) + \sum_{n=1}^N \pi_n \log\sigma_n + h(\pi).
\label{e:UpperboundK}
\end{equation}
Also, the lower bound of the entropy given by
eq.~(\ref{e:LowerBound}) reduces to
\begin{equation}
H(p) \ge \H(p) - \sum_{n=1}^N \pi_n [\epsilon_n' +
(\log\pi_n^{-1} + 1) \epsilon_n] .
\label{e:LowerBoundK}
\end{equation}
Let us arrange the $\mu_n$ by increasing order and take $\sigma_n$
small with respect to
\begin{equation}
\label{e:mindist} d_n \triangleq \min (\mu_n -
\mu_{n-1},\mu_{n+1}-\mu_{n})\enspace.
\end{equation}
where $\mu_0 = -\infty$ and $\mu_{N+1} = \infty$ by convention.
Under this assumption, the density~(\ref{e:density}) is strongly
multimodal and $\Omega_n$ in the above Lemma can be taken to be
intervals centered at $\mu_n$ of length $d_n$:
\begin{equation}
\Omega_n \triangleq (\mu_n-d_n/2, \mu_n+d_n/2).
\label{e:Omega_n}
\end{equation}
Then simple calculations give
\begin{eqnarray}
\left \{
\begin{array}{lll}
\epsilon_n &{}={}& 1 - \int_{-d_n/(2\sigma_n)}^{d_n/(2\sigma_n)}
K(x) dx\enspace,
\label{e:epsilon_n}\\
&&\nonumber\\ \epsilon_n' &{}={}& H(K) - H_{d_n/\sigma_n}(K) +
\epsilon_n\log(\sup K), \end{array}\right.\label{e:epsilon_n'}
\end{eqnarray}
where $H_\alpha(K) \triangleq -\int_{-\alpha/2}^{\alpha/2}
K(x)\log K(x)dx$. It is clear that $\epsilon_n$ and $\epsilon_n'$
both tend to 0 as $d_n/\sigma_n \to \infty$. Thus one gets the
following corollary.

\begin{coroll}
\label{coroll}
Let $p$ be given by~(\ref{e:density}) with $K_n$ of
the form~(\ref{e:kernel}) and $\sup_x K(x) < \infty$. Then $H(p)$
is bounded above by $\H(p)$ and converges to this bound as $\min_n
(d_n/\sigma_n) \to \infty$, $d_n$ being defined
in~(\ref{e:mindist}).
\end{coroll}

\subsubsection{Explicit calculation in the Gaussian case}
\label{GaussianCase}

Let us focus on the $K(x) = \Phi(x)$ case where $\Phi(x)$ denotes
the standard Gaussian density: $\Phi(x) = (1/\sqrt{2\pi})
\e^{-x^2/2}$.

The upper and lower bounds of $H(p)$ are given
by~(\ref{e:UpperboundK}) and (\ref{e:LowerBoundK}) with $H(\Phi)$
instead of $H(K)$; $\epsilon_n$ and $\epsilon_n'$ can now be
obtained explicitly :
\begin{eqnarray}
\left \{
\begin{array}{lll}
\epsilon_n =
\erfc\Big(\frac{d_n}{2\sqrt{2}\sigma_n}\Big),\nonumber\\
&&\nonumber\\
\epsilon_n' = H(\Phi) - H_{d_n/\sigma_n}(\Phi) -
\epsilon_n\log\sqrt{2\pi},
\end{array}
\right.
\end{eqnarray}
where $\erfc$ is the complementary error function defined as
$\erfc(x) = (2/\sqrt{\pi}) \int_x^\infty \exp(-z^2) dz$. By double
integration by parts and noting that $\int \erf(x) dx = x\erf(x) +
\exp(-x^2)/\sqrt{\pi}$ with $\erf(x)=1-\erfc(x)$, some algebraic
manipulations give
\[
\displaylines{
 H_{d_n/\sigma_n}(\Phi)=
\frac{1}{2}\erf\left(\frac{d_n}{2\sqrt{2}\sigma_n}\right)
\log(2\pi \e)\hfill{}\cr\hfill{} -
\frac{d_n}{2\sqrt{2\pi}\sigma_n}
\e^{-d_n^2/(8\sigma_n^2)}\enspace. }
\]
One can see that $H_{d_n/\sigma_n}(\Phi) \to
H(\Phi)=\log\sqrt{2\pi \e}$ as $d_n/\sigma_n\rightarrow\infty$,
as it should be. Finally:
\begin{eqnarray}
\left \{
\begin{array}{lll}
\epsilon_n &{}={}&
\displaystyle\erfc\Big(\frac{d_n}{2\sqrt{2}\sigma_n}\Big)
\\&&\nonumber\\
\epsilon_n'&{}={}& \displaystyle
\frac{1}{2}\erfc\Big(\frac{d_n}{2\sqrt{2}\sigma_n}\Big) +
\frac{d_n}{2\sqrt{2\pi}\sigma_n} \e^{-[d_n/(2\sqrt2\sigma_n)]^2}
\end{array}
\right. \label{e:EpsGauss}.\qquad
\end{eqnarray}

\begin{example}
\label{Ex:EntropApp3modPDF} To illustrate Corollary~\ref{coroll},
Fig.~\ref{Fig_EvolBound} plots the entropy of a trimodal variable
$Y$ with density $p$ as in (\ref{e:density}) with $K_n$ given by
(\ref{e:kernel}), $\sigma_n = \sigma$ (for the ease of
illustration), $K = \Phi$, $\mu = [0,5,10]$ and $\pi =
[1/4,1/2,1/4]$. Such variable can be represented as $Y = U +
\sigma Z$ where $U$ is a discrete random variable taking values in
$\{0,5,10\}$ with probabilities $1/4,1/2,1/4$ and $Z$ is a
standard Gaussian variable independent from $U$. The upper and
lower bounds of the entropy are computed as in Lemma 1 with the
above expressions for $\epsilon_n,~\epsilon_n'$, and plotted on
the same figure. One can see that the lower the $\sigma$, the
better the approximation of $H(Y)$ by its upper and lower bounds.
On the contrary, when $\sigma$ increases, the difference between
the entropy and its bounds tend to increase, which seems natural.
These differences however can be seen to tend towards a constant
for $\sigma \to \infty$. This can be explained as follows. When
$\sigma$ is large, $p$ is no longer multimodal and tends to the
Gaussian density of variance $\sigma^2$. Thus $H(Y)$ grows with
$\sigma$ as $\log\sigma$. On the other hand, the upper bound of
$\H(p)$ of $H(Y)$ also grows as $\log\sigma$. The same is true for
the lower bound of $H(Y)$ which equals $\H(p)-\sum_{n=1}^3 \pi_n
[\epsilon'_n + \epsilon_n(\log\pi_n^{-1} + 1)]$: the last term
tends to $h(\pi) + \frac{3}{2}$ as $\sigma \to \infty$ since for
fixed $d_n$, $\epsilon_n\rightarrow 1$ and $\epsilon_n'\rightarrow
1/2$ as $\sigma\rightarrow \infty$.

\begin{figure}
\centering
\includegraphics[width=8.5truecm]{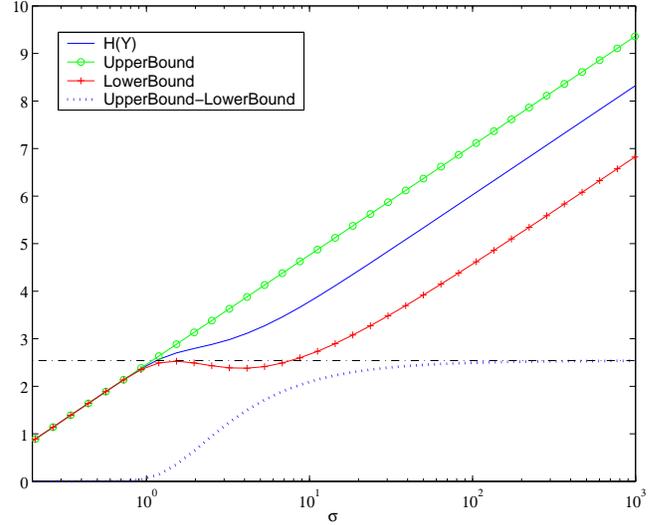}
\caption{Illustration of Example \ref{Ex:EntropApp3modPDF}:
Evolution of $H(Y)$ and its bounds versus $\sigma$, where
$Y=U+\sigma Z$, $U$ is a discrete random variable taking values in
$\{0,5,10\}$ with probabilities $\pi=[1/4,1/2,1/4]$ and $Z$ is a
standard Gaussian variable independent from $U$. The lower bound
converges to the upper bound as $\sigma\rightarrow0$ and the
difference between upper and lower bounds tends to $3/2+h(\pi)$ as
$\sigma\rightarrow\infty$ (note that the horizontal axis scale is
logarithmic).} \label{Fig_EvolBound}
\end{figure}
\end{example}

\subsection{Entropy bounds and decision theory}
\label{s:decision}

The entropy estimator given in eq. \eqref{Eq:ApproxSumEntrop} has
actually close connections with decision problems, and a tighter
upper bound for $H(p)$ can be found in this framework.
Assume we have a $N$-class classification problem consisting in
finding the class label $C$ of an observation $y_n$, knowing the
densities and the priors of the classes. In such kinds of
classification problems, one is often interested in quantifying
the Bayes' probability of error $P(e)$. In our context, each of
the pdf mode $K_n$ represents the density of a given class $c_n$,
i.e. the conditional density of $Y$ given $C = c_n$ is $K_n$.
Furthermore, $\pi_n$ is the a priori probability of $c_n$ : $P(C =
c_n) = \pi_n$, and $p$ is the density of $Y$, which can thus be seen as a ``mixture density''. 
Defining $h(C) =- \sum_{n=1}^N P(C = c_n) \log P(C = c_n)$, it can
be shown \cite{Hell70},\cite{Lin91} that
\begin{eqnarray}
\nonumber P(e) &\le& \frac{1}{2} h(C|Y) = \frac{1}{2} [H(Y|C) +
h(C) - H(Y)]
\\
\label{e:DecTheory} &=&  \frac{1}{2} \bigg[ \sum_{n=1}^N \pi_n
H(K_n) + h(\pi) - H(Y)\bigg]
\end{eqnarray}
where $H(Y|C) \triangleq \E_C[H(Y|C = c_i)]$, which shows that
half the difference between the $\H(p)$ and $H(p)$ is precisely an
upper bound of Bayes' probability of error $P(e) \triangleq \E_Y[1
- \max_i p(c_i|y)]$. The error vanishes when
the modes have no overlap (the classes are separable, i.e.
disjoint).

Clearly, ${\cal H}(p) - 2P(e)$ is a tighter upper bound of $H(p)$
than ${\cal H}(p)$ as $P(e) \ge 0$.
On the other hand, it can be proved that ${\cal H}(p) - 2\sqrt{(N
- 1)P(e)}$ is a lower bound for $H(p)$ \cite{Lin91}. However, the
lower bound in Lemma 1 is tighter when $\sigma$ is small enough.
Both bounds in this lemma are easier to deal with in more general
theoretical developments, are more related to the multimodality of
$p(y)$ and suffice for our purposes. Therefore, in the following
theoretical developments, the last pair of bounds shall be used.

\section{Mixing local minima in multimodal BSS}
\label{s:MixLocalMin}

Based on the results derived in Section \ref{Sec:EntropBounds}, it
will be shown that mixing local minima of the entropy exist in the
context of the blind separation of multimodal sources with
Gaussian modes if the mode standard deviations $\sigma_n$ are
\textit{small enough}.

We are interested in the (mixing) local minima of $H(\w\S)$ on the
unit sphere ${\cal S} = \{\w: \|\w\| = 1\}$ of $\mathbb{R}^K$. We
shall assume that the sources have a pdf of the
form~(\ref{e:density}), with $K_n$ being Gaussian with identical
variance $\sigma^2$ (but with distinct means). Thus, as in example
\ref{Ex:EntropApp3modPDF}, we may represent $S_k$ as $U_k + \sigma
Z_k$ where $U_k$ is a discrete random variable and $Z_k$ is a
standard Gaussian variable independent from $U_k$. Further, $(U_1,
Z_1), \dots, (U_K, Z_K)$ are assumed to be independent so that the
sources are independent as required. From this representation,
$\w\S = \w\U + \sigma Z$ where $\U$ is the column vector with
components $U_k$ and $Z$ is again a standard Gaussian variable
(since any linear combination of independent Gaussian variables is
a Gaussian variable and $\sum_{k=1}^K w_k Z_k$ has zero mean and
unit variance). Since $\w\U$ is clearly a discrete random
variable, $\w\S$ also has a multimodal distribution of the
form~(\ref{e:density}) with $K_n$ again the Gaussian density with
variance $\sigma^2$. Note that the number of modes is the number
of distinct values $\w\U$ can have and the mode centers (the means
of the $K_n$) are these values; they depend of $\w$. However, as
long as $\sigma$ is small enough with respect to the distances
$d_n$ defined in~(\ref{e:mindist}) the
approximation~(\ref{Eq:ApproxSumEntrop}) of the entropy is
justified. Thus, we are led to the approximation $H(\w\S) \approx
h(\w\U) + \log\sigma + H(\Phi)$, where $h(\w\U)$ denotes abusively
the entropy of the discrete random variable $\w\U$ (the entropy of
a discrete random variable $U$ with probability vector $\pi$ is
noted either $h(U)$ or $h(\pi)$).

The above approximation suggests that there is a relationship
between the local minimum points of $H(\w\S)$ and those of
$h(\w\U)$. Therefore, we shall first focus on the local minimum
points of the entropy of $\w\U$ before analyzing those of
$H(\w\S)$.

\subsection{Local minimum points of $h(\w\U)$}
\label{s:MixMinDiscr} The function $h(\w\U)$ does not depend on
the values that $\w\U$ can take but only on the associated
probabilities; these probabilities remain constant as $\w$ changes
unless the number of distinct values that $\w\U$ can take varies.
Such number would decrease when an equality $\w\u = \w\u'$ is
attained for some distinct column vectors $\u$ and $\u'$ in the
set of possible values of $\U$. A deeper analysis yields the
following result, which is helpful to find the local minimum point
of $h(\w\U)$.

\begin{lemma}
\label{lemma2} Let $\U$ be a discrete random vector in
$\mathbb{R}^K$ and $\cal U$ be the set of distinct values it can
take. Assume that there exists $r \ge 1$ disjoint subsets ${\cal
U}_1, \dots, {\cal U}_r$ of $\cal U$ each containing at least 2
elements, such that the linear subspace $V$ spanned by the vectors
$\u - \u_1, \u \in {\cal U}_1 \setminus \{\u_1\}, \dots, \u -
\u_r, \u \in {\cal U}_r \setminus \{\u_r\}$, $\u_1, \dots, \u_r$
being arbitrary elements of ${\cal U}_1, \dots, {\cal U}_r$, is of
dimension $K-1$. (Note that $V$ does not depend on the choice of
$\u_1, \dots, \u_r$, since $\u - \u_j' = (\u - \u_j) - (\u_j' -
\u_j)$ for any other $\u_j' \in {\cal U}_j$.) Then for $\w^* \in
\cal S$ and orthogonal to $V$, there exists a neighborhood $\cal
W$ of $\w^*$ in $\cal S$ and $\alpha > 0$ such that $h(\w\U) \ge
h(\w^*\U) + \alpha$ for all $\w \in {\cal W} \setminus \{\w^*\}$.
In the case $K=2$, one has a stronger result that $h(\w\U) = h(\U)
> h(\w^*\U)$ for all $\w \in {\cal W} \setminus \{\w^*\}$.

\end{lemma}

The proof is given in Appendix II.

\begin{example}
\label{ex:Cercle} An illustration of Lemma~\ref{lemma2} in the
$K=2$ case (again for clarity) is provided in
Fig.\ref{Fig_IllustCircle}. We note $\U=[U_1,U_2]^\T$ where the
discrete variables $U_1$ and $U_2$ take the values $-\sqrt{1.03}+
2.5, \sqrt{1.03}+ 2.5$ with probabilities and $.5,.5$ and the
values $-1.2,-.4, 2$ with probabilities $1/2,3/8,1/8$,
respectively. They are chosen to have the same variance, as we
need that the $S_k = U_k + \sigma Z_k$, $k = 1, 2$, have the same
variance. But their mean can be arbitrary since $H(\w\S)$ does not
depend on them. In this $K=2$ example, each line that links two
distinct points $\u,\u'\in{\cal U}$ span a one dimensional linear
subspace, which constitutes a possible subspace $V$, as stated in
Lemma~\ref{lemma2}. There are thus many possibilities for $V$,
each corresponding to a specific vector $\w^*$.

Two simple possibilities for $V$ are the subspaces with direction
given by $[0,1]^\T$ and $[1,0]^\T$. In the first case, the subsets
${\cal U}_i$ are built by grouping the points of ${\cal U}$ laying
on a same vertical dashed line. There are two such subsets ($r =
2$) consisting of $\u \in \cal U$ with first component equal to
$-\sqrt{1.03}+2.5$ and $\sqrt{1.03}+2.5$, respectively. In the
second case, the subsets ${\cal U}_i$ are built by grouping the
points of ${\cal U}$ laying on a same horizontal dashed line.
There are three such subsets ($r = 3$) consisting of $\u \in \cal
U$ with second component equal to $-1.2$, $-.4$ and 2,
respectively.

There also exist other subspaces $V$, corresponding to ``diagonal
lines'' (i.e. to solid lines in Fig.\ref{Fig_IllustCircle}). This
last kind of one-dimensional linear subspace $V$ correspond to
directions given by two-dimensional vectors $\w^*$ with two
non-zero elements.

On the plot, the points on the half circle correspond to the
vectors $\w^*$ of the Lemma; each  $\w^*$ is orthogonal to a line
joining a pair of distinct points in $\cal U$, $\cal U$ being the
set of all possible values of $[U_1, U_2]^\T$. The points of $\cal
U$ are displayed in the plot together with their probabilities.
The entropies $h(\w\U)$ are also given in the plot; one can see
that they are lower for $\w = \w^*$ than for other points $\w$.

\begin{figure}
\centering
\includegraphics[width=8.5truecm]{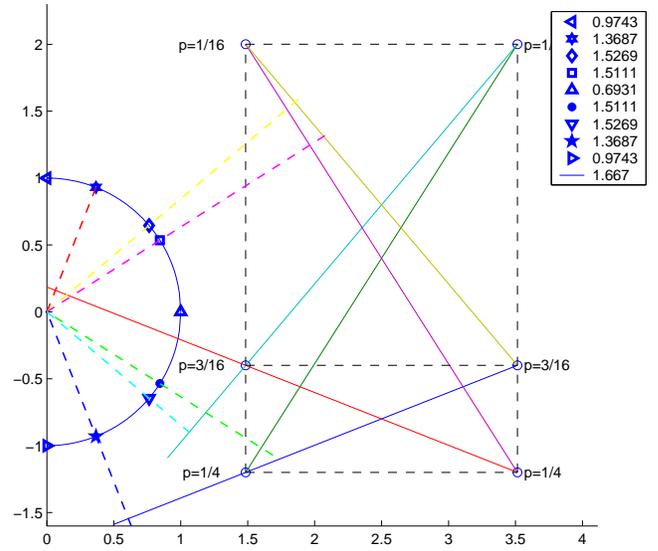}
\caption{Example 5: illustration of Lemma 2. The discrete random
variables $U_1$ and $U_2$ take values in $\{-\sqrt{1.03}+ 2.5,
\sqrt{1.03}+ 2.5\}$ and $\{-1.2,-.4, 2\}$ with probabilities $[.5
.5]$ and $[1/2,3/8,1/8]$, respectively.}\label{Fig_IllustCircle}
\end{figure}
\end{example}

The above Lemma only provides a mean to find a local minimum point
of the function $h(\w\U)$, but does not prove the existence of
such a point, since the existence of $V$ was only {\em assumed} in
the Lemma. Nevertheless, in the case where the components of $\U$
are independent and can take at least 2 distinct values, subset
${\cal U}_i$ ensuring the existence of $V$ can be built as
follows. Let $j$ be any index in $\{1, \dots, K\}$ and
$\lambda_{j,1}, \dots, \lambda_{j,r_j}$ be the possible value of
$U_j$, the $j$-th component of $\U$. One can take ${\cal U}_i, 1
\le i \le r_j$ to be the set of $\u \in \cal U$ such that its
$j$-th components equal $\lambda_{j,i}$. Then it is clear that the
corresponding subspace $V$ consists of all vectors orthogonal to
the $j$-th row of the identity matrix (hence $V$ is of dimension
$K-1$) and that the associated vector $\w^*$ is simply this row or
it opposite. By Lemma~\ref{lemma2}, this point $\w^*$ would be a
local minimum point of $h(\w\U)$. But, as explained above, it is a
non mixing point while we are interested in the mixing point, i.e.
not proportional to a row of the identity matrix. However, the
above construction can be extended by looking for a set of $K$
vectors $\u_1, \dots, \u_K$ in $\cal U$, such that the vectors
$\u_i - \u_j, 1 \le i < j \le K$ span any linear subspace of
dimension $K-1$ of $\mathbb{R}^K$. If such a set can be found,
then $V$ is simply this linear subspace by taking ${\cal U}_1 =
\{\u_1, \dots, \u_K\}$ and $r=1$. In addition, if $\u_1, \dots,
\u_K$ do not all have the same $j$-th component, for some $j$,
then the corresponding $\w^*$ is a mixing local minimum point. In
view of the fact that there are at least $2^K$ points in $\cal U$
to choose from for the $\u_i$ and that the last construction
procedure meant not find all local minimum points of $h(\w\U)$,
chance is that there exists both non-mixing and mixing local
minimum points of $h(\w\U)$. In the $K=2$ case this is really the
case: it suffices to take two distinct points $\u_1$ and $\u_2$ in
$\cal U$, then by the above Lemma, the vector $\w^*$ orthogonal to
$\u_1 - \u_2$ is a local minimum point of $h(\w\U)$. If one choose
$\u_1$ and $\u_2$ such that both components of $\u_1 - \u_2$ are
non zero, the associated orthogonal vector $\w^*$ is not
proportional to any row of the identity matrix; it is a mixing
local minimum point of $h(\w\U)$. Note that in the particular
$K=2$ case, the aforementioned method identifies all local minimum
points of $h(\w\U)$. Indeed, for any $\w \in \cal S$, either there
exists a pair of distinct vectors $\u_1, \u_2$ in $\cal U$ such
that $\w(\u_1 - \u_2) = 0$ or there exists no such pair. In the
first case $\w$ is a local minimum point and in the second case
one has $h(\w\U) = h(\U)$. Since there is only a finite number of
the differences $\u_1 - \u_2$, for distinct $\u_1, \u_2$ in $\cal
U$, there can be only a finite number of local minimum points of
$h(\w\U)$, and for all other points $h(\w\U)$ take the maximum
value $h(\U)$.

\subsection{Local minimum points of $H(\w\S)$}

This subsection shows that the local minima points of $H(\w\S)$
can be related to those of $h(\w\U)$.

\begin{lemma}
\label{lemma3} Define $S_i$, $i = 1,\cdots, K$, as $S_i = U_i +
\sigma Z_i$ described at the beginning of
subsection~\ref{s:MixLocalMin} and $\w^*$ be a vector satisfying
the assumption of Lemma~\ref{lemma2} ($\U$ being the vector with
component $U_i$). Then for $\sigma$ sufficiently small $H(\w\S)$
admits a local minimum point converging to $\w^*$ as $\sigma \to
0$.
\end{lemma}

The proof of this Lemma is relegated to the Appendix.

\begin{example}
\label{example6}
Thanks to the entropy approximator, we shall
illustrate the existence of the local minima of $H(\w\S)$ in the
following $K = 2$ example, so that vectors $\w$ satisfying
$||\w||=1$ can be written as $[\sin\theta, \cos\theta]$. We take
$S_1 = U_{\pi/2}+\sigma Z_1$ and $S_2 = U_0+\sigma Z_2$, where
$U_{0},U_{\pi/2}$ are independent discrete random variables taking
the values $-2\sqrt{3}/3,\sqrt{3}/2$ with probabilities $1/3,2/3$
and $-\sqrt{2},\sqrt{2}/2$ with probabilities $3/7,4/7$,
respectively, and $Z_1$, $Z_2$ are standard Gaussian variables.
The parameter $\sigma$ is set to 0.1. Thus $Y_{\theta} = \w\S$ can
be represented as $U_{\theta} + \sigma Z$ where $U_{\theta} =
\sin\theta U_{\pi/2} + \cos\theta U_0$ and $Z$ is a standard
Gaussian variable independent from $U_\theta$.
Figure~\ref{Fig_OutPDF} plots the pdf of $Y_\theta$ for various
angles $\theta$. It can be seen that the modality (i.e. the number
of modes) changes with $\theta$.
Fig. \ref{Fig_Entrop2x2_d(n)} shows the entropy of $Y_\theta$
together with its upper and lower bounds, for $\theta\in[0,\pi]$.
In addition to non-mixing local minima at
$\theta\in\{p\pi/2|p\in\mathbb{Z}\}$, mixing local minima exist
when $\w(\u_1-\u_2) = 0$, where $\u_1 = [-2\sqrt{3}/3,
\sqrt{2}/2]^\T, \u_2 = [\sqrt{3}/2,-\sqrt{2}]^\T$, i.e. when
$|\tan(\theta)|=.9526$, or $\theta\in\{(0.2423+p)\pi,
(0.7577+p)\pi|p\in\mathbb{Z}\}$. One can observe that the upper
bound is a constant function except for a finite number of angles
for which we observe negative peaks (see Lemma~\ref{lemma2}). For
these angles the pdf is strongly multimodal, and the upper and
lower bounds are very close, though not clearly visible on the
figure. This results from a discontinuity of the lower bound at
these angles, due to the superimposition of several modes at these
angles.

\begin{figure}
\centering
\includegraphics[width=8.5truecm]{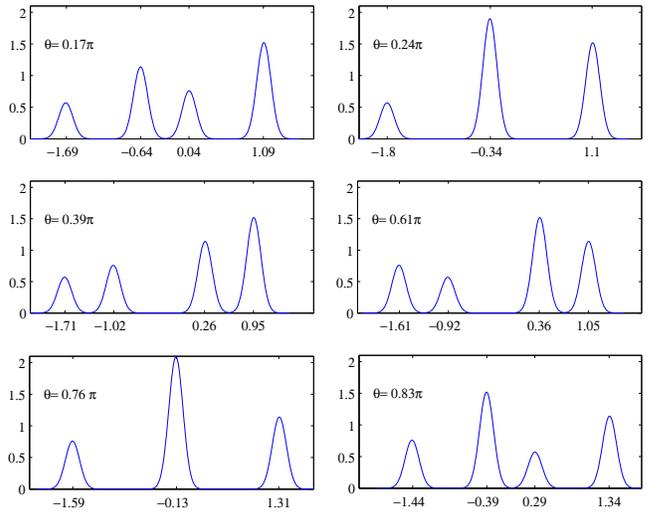}
\caption{Example 6: probability density function of $\w\S$ for
various angles $\theta$.} \label{Fig_OutPDF}
\end{figure}

\begin{figure}
\centering
\includegraphics[width=8.5truecm]{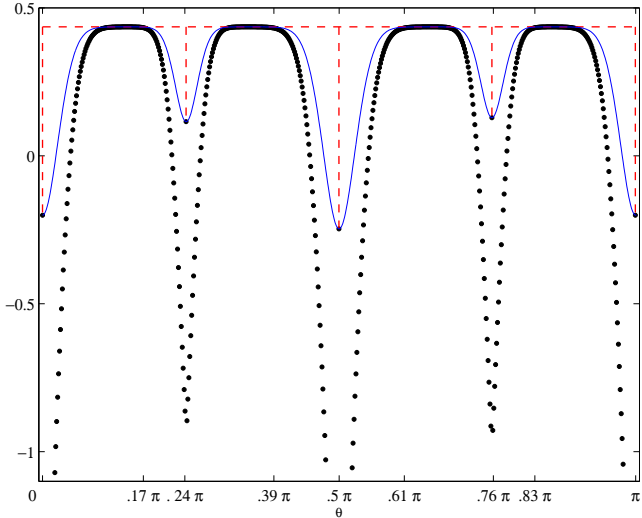}
\caption{Example 6: Upperbound (dashed line), lower bound (dots)
and entropy estimation of $Y_\theta$ using finite Riemannian sum
(solid). It can be seen that the upper and lower bounds of the
entropy converge to each other when the density becomes strongly
multimodal (see the corresponding plots in Fig.
\ref{Fig_OutPDF}).} \label{Fig_Entrop2x2_d(n)}
\end{figure}
\end{example}

\section{Complementary observations}

This section provides two observations that can be drawn regarding
the impact of the \emph{mode variance} $\sigma^2$ on the existence
of local minima and the symmetry of the entropy with respect to
$\theta$.

\subsection{Impact of ``mode variance'' $\sigma^2$}
\begin{figure}
\centering
\includegraphics[width=8.5truecm]{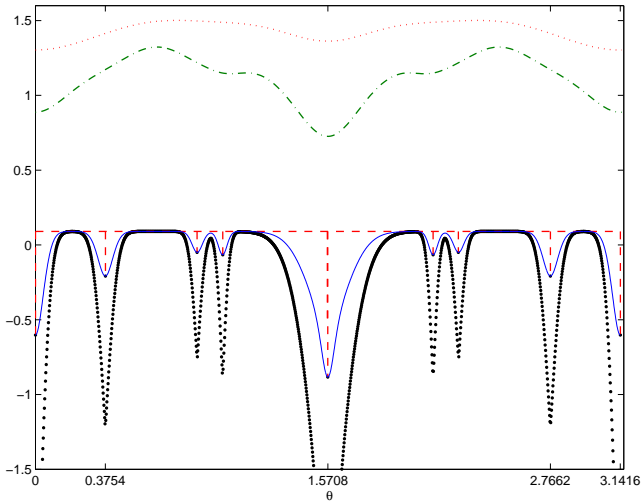}
\caption{Entropy of $\w\S$ (estimated using finite Riemannian sum)
versus $\theta$ for $S_1=U_1+\sigma Z_1$, $S_2=U_2+\sigma Z_2$,
where $U_1$ and $U_2$ are taken from example~\ref{ex:Cercle} (and
Fig. \ref{Fig_IllustCircle}) and the four random variables are all
independent. The parameter $\sigma$ is set to $.05$ (solid), $.25$
(dashed-dotted) and $.5$ (dotted). The upper and lower bounds have
been added for the $\sigma=.05$ case only, for visibility
purposes. It can be seen that the upper and lower bounds of the
entropy converge to each other when the density becomes strongly
multimodal.} \label{Fig_Isspa}
\end{figure}

In the example of Fig.~\ref{Fig_Isspa} the discrete variables
$U_1$ and $U_2$ in the expression of $S_1$ and $S_2$ are taken as
in Example~\ref{ex:Cercle}. One can observe that the mixing minima
of the entropy tends to disappear when the mode variance
increases. This is a direct consequence of the fact that the mode
overlaps increase. When $\sigma$ increases, the source densities
become more and more Gaussian and the $H(\w\S)$ vs $\theta$ curve
tends to be more and more flat, approaching the constant function
$\log\sqrt{2\pi\e}+\log\sigma$. The upper and lower bounds have
only been plotted for the $\sigma=.05$, for visibility purposes.
Again, at angles corresponding to the upper bound negative peaks,
the error bound is very tight, as explained in Example
\ref{example6}.

\subsection{Note on symmetry of $H(\w_\theta\S)$}
In the above graphs plotting the entropy (and its bounds) versus
$\theta$, some symmetry can be observed. First, if we note
$\w_\theta=[\sin\theta \cos\theta]$, observe that
$H(\w_\theta\S)=H(\w_{\theta+\pi}\S)$ whatever are the source
pdfs; this is a direct consequence of the fact the the entropy is
not sign sensitive. Second, if one of the source densities is
symmetric, i.e. if it exists $\mu\in\mathbb{R}$ so that
$p_{S_j}(\mu-s)=p_{S_j}(\mu+s)$ for all $s \in \mathbb{R}$, then
$H(\w_\theta\S)=H(\w_{-\theta}\S)$. Third, if the two sources
share the same pdf, then $H(\w_\theta\S)=H(\w_{\pi/2-\theta}\S)$.
Finally, if the two sources can be expressed as in
Lemma~\ref{lemma3}, then the vectors $\w^*$ for which $h(\w^*\U) <
h(\U)$ (as obtained in Lemma~\ref{lemma2}) are symmetric in the
sense that their angles are pairwise opposite. This means that for
$\sigma$ small enough, if a local minimum of $H(\w_\theta\S)$
appears at $\theta^*$, then another local minimum point will exist
near $-\theta^*$ (and thus near $p\pi - \theta,\ \forall p \in
\mathbb{Z})$. The above symmetry property can be seen from
Figure~\ref{Fig_IllustCircle} and can be proved formally as
follows. From Lemma~\ref{lemma2}, $\w^*$ must be orthogonal to
$\u_1 - \u_2$ for some pair of distinct vectors in the set of all
possible values of $\U$. Define $\u_i^\dagger$ ($i = 1,2$) to be
the vector with first coordinate the same as that of $\u_{3-i}$
and second coordinate the same as that of $\u_i$. Then it can be
seen that the vector orthogonal to $\u_1^\dagger - \u_2^\dagger$
has an angle opposite to the angle of $\w^*$, yielding the desired
result.

\section{Conclusion}

In this paper, new results regarding both non-mixing and mixing
entropy local minima have been derived in the context of the blind
separation of $K$ sources. First, it is shown that a local entropy
minimum exists when the output is proportional to one of the
non-Gaussian source. Second, it is shown that mixing entropy
minima may exist when the source densities are strongly multimodal
(i.e. multimodal with sufficiently small overlap); therefore,
spurious BSS solutions can be obtained when minimizing this
entropic criterion. Some attention must be paid to the obtained
solutions when they are found by adaptive gradient minimization.

To prove the existence of mixing entropy minima, a theoretical
framework using an entropy approximator and its associated error
bounds has been provided. Even if this approximator is considered
here in the context of blind source separation, its use can be
extended to other applications involving entropy estimation.

\section*{Acknowledgment}
The authors are grateful to the anonymous referees for their
constructive remarks that have contributed to improve the quality
paper. More specifically, the authors are indebted to reviewer B
for having provided a simple way for proving inequality
\eqref{e:UpperBound}.

\bibliography{Biblio}

\begin{thebibliography}{10}
\providecommand{\url}[1]{#1}
\def\UrlFont{\rmfamily}
\providecommand{\newblock}{\relax}
\providecommand{\bibinfo}[2]{#2}
\providecommand\BIBentrySTDinterwordspacing{\spaceskip=0pt\relax}
\providecommand\BIBentryALTinterwordstretchfactor{4}
\providecommand\BIBentryALTinterwordspacing{\spaceskip=\fontdimen2\font plus
\BIBentryALTinterwordstretchfactor\fontdimen3\font minus
  \fontdimen4\font\relax}
\providecommand\BIBforeignlanguage[2]{{%
\expandafter\ifx\csname l@#1\endcsname\relax
\typeout{** WARNING: IEEEtran.bst: No hyphenation pattern has been}%
\typeout{** loaded for the language `#1'. Using the pattern for}%
\typeout{** the default language instead.}%
\else
\language=\csname l@#1\endcsname
\fi
#2}}

\bibitem{cit_Como94}
P.~Comon, ``Independent component analysis, a new concept?'' \emph{Signal
  Processing}, vol.~36, no.~3, pp. 287--314, 1994.

\bibitem{cit_ICABook}
A.~Hyv\"{a}rinen, J.~Karhunen, and E.~Oja, \emph{Independent component
  analysis}.\hskip 1em plus 0.5em minus 0.4em\relax New York: John Willey and
  Sons, Inc., 2001.

\bibitem{cit_BSS}
S.~Haykin, Ed., \emph{Unsupervised Adaptive Filtering vol.1 : Blind Source
  Separation}.\hskip 1em plus 0.5em minus 0.4em\relax New York: John Willey and
  Sons, Inc., 2000.

\bibitem{cit_CichokiBook}
A.~Cichoki and S.-I. Amari, \emph{Adaptive blind signal and image
  processing}.\hskip 1em plus 0.5em minus 0.4em\relax England: John Willey and
  Sons, Inc., 2002.

\bibitem{cit_MI}
T.~M. Cover and J.~A. Thomas, \emph{Elements of information theory}.\hskip 1em
  plus 0.5em minus 0.4em\relax Wiley and sons, 1991.

\bibitem{cit_GrayIT}
R.~M. Gray, \emph{Entropy and Information Theory}.\hskip 1em plus 0.5em minus
  0.4em\relax Springer-Verlag, New York, 1991.

\bibitem{cit_PhamIEEE_IT}
D.-T. Pham, ``Mutual information approach to blind separation of stationary
  sources,'' \emph{{IEEE} {T}rans. {I}nform. {T}heory}, vol.~48, no.~7, pp.
  1935--1946, 2002.

\bibitem{cit_Deflation}
N.~Delfosse and P.~Loubaton, ``Adaptibe blind separation of sources: A
  deflation approach,'' \emph{Signal Processing}, vol.~45, pp. 59--83, 1995.

\bibitem{cit_FastICA}
A.~Hyv\"{a}rinen, ``Fast and robust fixed-point algorithms for independent
  component analysis,'' \emph{{IEEE} Trans. Neural Networks}, vol.~10, no.~3,
  pp. 626--634, 1999.

\bibitem{cit_Cruces-IEEE-NN}
S.~Cruces, A.~Cichocki, and S.~Amari, ``From blind signal extraction to blind
  instantaneous signal separation: criteria, algorithms and stability,''
  \emph{{IEEE} {T}rans. {N}eural {N}etworks}, vol.~15, no.~4, pp. 859--873,
  July 2004.

\bibitem{cit_PhamIEESP_ICA}
D.-T. Pham, ``Blind separation of instantaneous mixture of sources via an
  independent component analysis,'' \emph{{IEEE} {T}rans. {S}ignal
  {P}rocessing}, vol.~44, no.~11, pp. 2768--2779, 1996.

\bibitem{cit_Gray}
R.~M. Gray and L.~D. Davisson, \emph{An Introduction to Statistical Signal
  Processing}.\hskip 1em plus 0.5em minus 0.4em\relax Cambridge University
  Press, 2004.

\bibitem{cit_Dembo}
A.~Dembo, T.~M. Cover, and J.~A. Thomas, ``Information theoretic
  inequalities,'' \emph{IEEE Trans. Inform. Theory}, vol.~37, no.~6, pp.
  1501--1518, 1991.

\bibitem{cit_vrins_SP_04}
F.~Vrins and M.~Verleysen, ``On the entropy minimization of a linear mixture of
  variables for source separation,'' \emph{{S}ignal {P}rocessing}, vol.~85,
  no.~5.

\bibitem{cit_vrins_IEEESPL_05a}
------, ``Information theoretic vs cumulant-based contrasts for multimodal
  source separation,'' \emph{{IEEE} {S}ignal {P}rocessing {L}ett.}, vol.~12,
  no.~3, pp. 190--193, 2005.

\bibitem{cit_learned-miller_04}
E.~G. Learned-Miller and J.~W. {Fisher III}, ``{ICA} using spacings estimates
  of entropy,'' \emph{Journal of Machine Learning Research}, vol.~4, pp.
  1271--1295, 2003.

\bibitem{cit_vrins_Eusipco05_a}
F.~Vrins, J.~Lee, and M.~Verleysen, ``Can we always trust entropy minima in the
  ica context ?'' in \emph{Eur. Signal Processing Conf. ({EUSIPCO}'05)},
  Antalya (Turkey), pp. cr1107.1--cr1107.14.

\bibitem{cit_vrins_IEEESPL_05b}
D.-T. Pham and F.~Vrins, ``Local minima of information-theoretic criteria in
  blind source separation,'' \emph{{IEEE} {S}ignal {P}rocessing {L}ett.},
  vol.~12, no.~11, pp. 788--791, 2005.

\bibitem{cit_PhamVrins_ISPPA05}
D.-T. Pham, F.~Vrins, and M.~Verleysen, ``Spurious entropy minima for
  multimodal source separation,'' in \emph{Int. Symp. on Signal Processing and
  Applications ({ISSPA}'05)}, Sidney (Australia), pp. 37--40.

\bibitem{cit_PhamEntrop}
D.-T. Pham, ``Entropy of a variable slightly contaminated with another,''
  \emph{{IEEE} {S}ignal {P}rocessing {L}ett.}, vol.~12, no.~7, pp. 536--539,
  2005.

\bibitem{Silverman1986}
B.~W. Silverman, \emph{Density Estimation}.\hskip 1em plus 0.5em minus
  0.4em\relax Chapman, Hall/CRC (London), 1986.

\bibitem{Scott92}
D.~W. Scott, \emph{Multivariate Density Estimation: theory, practice and
  visualization}.\hskip 1em plus 0.5em minus 0.4em\relax John Wiley and Sons
  (New York), 1992.

\bibitem{Hell70}
J.~R. M.~Hellman, ``Probability of error, equivocation, and the chernoff
  bound,'' \emph{{IEEE} Trans. Inform. Theory}, vol.~16, no.~4.

\bibitem{Lin91}
J.~Lin, ``Divergence measures based on the shannon entropy,'' \emph{{IEEE}
  Trans. Inform. Theory}, vol.~37, no.~1.

\end{thebibliography}
\bibliographystyle{IEEEtran}

\appendices
\section{Proofs of Lemmas}

\noindent{\bf Proof of Lemma~\ref{lemma1}} We have
from~(\ref{e:EntropPy}) that $H(Y) = \sum_{n=1}^N \pi_n H_n$ where
\begin{equation}
H_n \triangleq - \int K_n(y) \log\Big[\sum_{m=1}^N \pi_m
K_m(y)\Big] dy. \label{e:Hn}
\end{equation}
Since all $K_m \ge 0$, the last right hand side is bounded above
by $-\int K_n(y)\log[\pi_n K_n(y)]\,dy = H(K_n) - \log\pi_n$,
yielding the inequality \eqref{e:UpperBound}.

A more elegant derivation of this inequality can be obtained from
the entropy properties. Indeed, the density given in
\eqref{e:density} can be interpreted as the marginal density of an
augmented model $(Y,U)$ where $U$ is a discrete variable with $N$
values $u_1, \dots, u_n$
with probabilities $\pi_1, \dots, \pi_n$ and $Y$ has a conditional
density given $U = u_n$ equal to $K_n$. The joint entropy $H(Y,U)$
of (the ``continuous-discrete'' pair of random variables) $Y, U$
equals $H(Y|U) + h(U)$ where $h(U) = h(\pi)$ is the discrete
entropy of $U$ and $H(Y|U) = \sum_{n=1}^N \pi_n H(K_n)$ is the
conditional entropy of $Y$ given $U$. But $H(Y,U) = h(U|Y) + H(Y)$
(where $h(U|Y)$ is the conditional entropy of $U$ given $Y$) and
thus $\H(p) - H(p)$ equals $h(U|Y)$ which is always nonnegative
because $U$ is a discrete variable.

Yet another way to prove the above inequality is to exploit its
connection to the decision problem discussed in
Section~\ref{s:decision}. Indeed, equation~(\ref{e:DecTheory})
yields immediately $\H(p) - H(p) \ge P(e) \ge 0$.

To prove the second result, noting that $\log(1+x) \le x$, the
term $\log[\sum_{m=1}^N \pi_m K_m(y)]$ can be bounded above by
\begin{eqnarray}
\left \{
\begin{array}{ll}
\displaystyle
\log[\pi_n K_n(y)] +
\sum_{1 \le m \le N, m \ne n} \frac{\pi_m K_m(y)}{\pi_n K_n(y)} &
\hbox{ if } y \in \Omega_n
\\
\log(\max_{1 \le m \le N} \sup K_m) & \hbox{otherwise}\enspace.
\end{array}
\right.
\end{eqnarray}
Therefore, with 
\begin{equation}
H_n \triangleq - \int K_n(y) \log\Big[\sum_{m=1}^N \pi_m
K_m(y)\Big] dy. \label{e:Hn}
\end{equation}
one gets
\begin{eqnarray*}
H_n &{}\ge{}& - \int_{\Omega_n} K_n(y) \log[\pi_n K_n(y)] dy
\\ &&
- \sum_{1 \le m \le N, m \ne n} \frac{\pi_m}{\pi_n}
\int_{\Omega_n} K_m(y) dy
\\ &&
- \log (\max_{1 \le m \le N} \sup K_m) \epsilon_n
\end{eqnarray*}
But since $\Omega_1, \dots, \Omega_N$ are disjoint,
\[
\ifx\twocol\undefined
\sum_{n=1}^N \pi_n \sum_{1 \le m \le N, m
\ne n} \frac{\pi_m}{\pi_n} \int_{\Omega_n} K_m(y) dy =
\sum_{m=1}^N \pi_m
\int_{\cup_{1 \le n \ne m \le N} \Omega_n} K_m(y) dy,
\else
\displaylines{
\sum_{n=1}^N \pi_n \sum_{1 \le m \le N, m
\ne n} \frac{\pi_m}{\pi_n} \int_{\Omega_n} K_m(y) dy ={}
\hfill\cr\hfill
\sum_{m=1}^N \pi_m
\int_{\cup_{1 \le n \ne m \le N} \Omega_n} K_m(y) dy,
}
\fi
\]
and $\cup_{1 \le n \ne m \le N} \Omega_n \subseteq \mathbb{R}
\setminus \Omega_m$. Therefore the right hand side of the above
equality is bounded above by $\sum_{m=1}^N \pi_m \epsilon_m$. It
follows that $H(p) = \sum_{n=1}^N \pi_n H_n$ is bounded below by
\[
\displaylines{ h(\pi) + \sum_{n=1}^N \pi_n H(K_n) + \sum_{n=1}^N
\pi_n \log(\pi_n \sup K_n) \epsilon_n - \sum_{n=1}^N \pi_n
\epsilon_n' \hfill\cr\hfill -\sum_{m=1}^N \pi_m \epsilon_m -
\sum_{n=1}^N \pi_n \log(\max_{1 \le m \le N} \sup K_m) \epsilon_n
}
\]

After some manipulations, the above expression reduces to the lower
bound for $\sum_{n=1}^N \pi_n H_n$ given in the Lemma \qed

\medskip
\noindent{\bf Proof of Lemma~\ref{lemma2}}

By construction, for each $j = 1, \dots, r$, $\w^*\u$ take the
same values for $\u \in {\cal U}_j$. On the other hand, by
grouping the vectors $\u \in \cal U$ which produce the same value
of $\w^*\u$ into subsets of $\cal U$, one gets a partition of
$\cal U$ into $r^* + 1$ subsets ${\cal U}_0^*, \dots, {\cal
U}_{r^*}^*$, such that each ${\cal U}_j^*, 1 \le j \le r^*$
contains at least two elements and $\w^*\u$ takes the same values
for $\u \in {\cal U}_j^*$ and the values associated with different
${\cal U}_j^*$ and the $\w^*\u, \u \in {\cal U}_0^*$, are all
distinct. Obviously $r^* \ge 1$ and each of the ${\cal U}_1,
\dots, {\cal U}_r$, must be contained in one of the ${\cal U}_1^*,
\dots, {\cal U}_{r^*}^*$. Therefore the space $V$ must be
contained in the space spanned by the vectors $\u - \u_j, \u \in
{\cal U}_j^* \setminus \{\u_j\},\ j = 1, \dots, r^*$, $\u_1,
\dots, \u_{r^*}$ being arbitrary elements of ${\cal U}_1^*, \dots,
{\cal U}_{r^*}^*$. But the last space is orthogonal to $\w^*$ by
construction and thus cannot have dimension greater than $K-1$,
hence it must coincide with $V$.

Putting $P(\u)$ for $P(\U = \u)$ for short and $P({\cal U}_j^*) =
\sum_{\u \in {\cal U}_j^*}P(\u)$, one has
\[
h(\w^*\U) = - \sum_{\u \in {\cal U}_0^*} P(\u) \log P(\u) -
\sum_{j=1}^{r^*} P({\cal U}_j^*) \log P({\cal U}_j^*) .
\]

For a given pair $\u, \u'$ of distinct vectors in $\cal U$, if
$\w^*(\u - \u') \ne 0$ then it remains so when $\w^*$ is changed
to $\w$ provided that the change is sufficiently small. But if
$\w^*(\u - \u') = 0$ then this equality may break however small
the change. In fact if $\w$ is not proportional to $\w^*$, it is
not orthogonal to $V$, hence $\w(\u - \u') \ne 0$ for at least one
pair $\u, \u'$ of distinct points in some ${\cal U}_j^*$, meaning
that $\w\u$ takes at least two distinct values in ${\cal U}_j^*$.
Thus there exists a neighborhood of $\cal W$ of $\w^*$ in $\cal S$
such that for all $\w \in \cal W \setminus \{\w^*\}$, each subset
${\cal U}_j^*$ be partitioned into subsets ${\cal U}_{j,k}(\w), k
= 1, \dots, n_j(\w)$ ($n_j(\w)$ can be 1) such that $\w\u$ takes
the same value on ${\cal U}_{j,k}(\w)$, and the values of $\w\u$
on the subsets ${\cal U}_{j,k}(\w)$ and on each points of ${\cal
U}_0^*$ are distinct. Further, there exists at least one index $i$
for which $n_i(\w) > 1$. For such an index
\[
\displaylines{ P({\cal U}_i^*) \log P({\cal U}_i^*) =
\sum_{k=1}^{n_i(\w)} P[{\cal U}_{i,k}(\w)] \log P[{\cal
U}_{i,k}(\w)] +{} \hfill\cr\hfill \sum_{k=1}^{n_i(\w)} P[{\cal
U}_{i,k}(\w)] \log \frac{P({\cal U}_i^*)}{P[{\cal
U}_{i,k}(\w)]}\enspace. }
\]
The last term can be seen to be a strictly positive number, as
$P({\cal U}_i^*) > P[{\cal U}_{i,k}(\w)]$ for $1 \le k \le
n_i(\w)$. Note that this term does not depend directly on $\w$ but
only indirectly via the set ${\cal U}_{j,k}(\w), k = 1, \dots,
n_j(\w), j = 1, \dots, r^*$, and there is only a finite number of
possible such sets. Therefore $h(\w\U) \ge h(\w^*\U) + \alpha$ for
some $\alpha > 0$ for all $\w \in \cal W$.

In the case $K=2$, the space $V$ reduces to a line and thus the
differences $\u - \u'$ for distinct $\u, \u'$ in ${\cal U}_j^*$,
for all $j$, are proportional to this line. Thus if $\w$ is not
proportional to $\w^*$, hence not orthogonal to this line, $\w\u$
take distinct values on each of the sets ${\cal U}_1^* \dots,
{\cal U}_{r^*}^*$, and if $\w$ is close enough to $\w^*$, these
values are also distinct for different sets and distinct from the
values of $\w\u$ on ${\cal U}_0^*$, which are distinct themselves.
Thus for such $\w$, $h(\w\U) = h(\U)$. \qed

\medskip
\noindent{\bf Proof of Lemma~\ref{lemma3}} The proof of this Lemma
is quite involve in the $K > 2$ case, therefore, we will first
give the proof for the $K=2$ case which is much simpler, and then
proceed by extending it to $K>2$. As already shown in the
beginning of section~\ref{s:MixLocalMin}, $\w\S = \w\U + \sigma Z$
where $Z$ is a standard Gaussian distribution. Thus, the density
of $\w\S$ is of the form~(\ref{e:density}) with $K_n(y) = \Phi[(y
- \mu_n)/\sigma]/\sigma$, $\mu_1, \dots, \mu_N$ being the possible
values of $h(\w\U)$ and $\Phi$ being the standard Gaussian
density. For $\w = \w^*$, one has by Lemma~\ref{lemma1},
\[
H(\w^*\S) \le h(\w^*\U) + H(\Phi) + \log\sigma.
\]

On the other hand, we have seen in the proof of Lemma~\ref{lemma2}
that for $\w$ in some neighborhood $\cal W$ of $\w^*$ and distinct
from $\w$, the $\w\u, \u \in \cal U$ ($\cal U$ denoting the set of
possible values of $\U$) are all distinct (in the $K=2$ case).
Thus the maps $\u \mapsto \w\u$ map different points $\u \in \cal
U$ to different $\mu_n$. However, when $\w$ approaches $\w^*$,
some of the $\mu_n$ tend to coincide and thus some of the $d_n$
defined in~(\ref{e:mindist}) approach zero. To avoid this we
restrict $\w$ to ${\cal W} \setminus {\cal W}'$ where ${\cal W}'$
is any open neighborhood of $\w^*$ strictly included in $\cal W$.
Then $\min_n d_n \ge d$ for all $\w \in {\cal W} \setminus {\cal
W}'$ for some $d > 0$ (which depends on ${\cal W}'$). Thus by
Corollary~\ref{coroll}, $H(\w\S)$ can be made arbitrarily close to
$h(\w\U) + H(\Phi) + \log\sigma$ for all $\w \in {\cal W}
\setminus {\cal W}'$ by taking $\sigma$ small enough. But $h(\w\U)
= h(\U) > h(\w^*\U)$, therefore $H(\w\S) > H(\w^*\S)$ for all $\w
\in {\cal W} \setminus {\cal W}'$, for $\sigma$ small enough.

One can always choose $\cal W$ to be a close set in $\cal S$;
hence it is compact. Since the function $\w \in {\cal W} \mapsto
H(\w\S)$ is continuous, it must admit a minimum, which by the
above result must be in ${\cal W}'$ and thus is not on the
boundary of $\cal W$. This shows that this minimum is a local
minimum. Finally, as one can choose ${\cal W}'$ arbitrarily small,
the above result shows that the above local minimum converges to
$\w^*$ as $\sigma \to 0$.

Consider now the case $K > 2$. The difficulty is that it is no
longer true that for $\w$ in some neighborhood $\cal W$ of $\w^*$
and distinct from $\w$, the $\w\u, \u \in \cal U$ are all
distinct. Indeed, by construction of $\w^*$, there exists $K-1$
pairs $(\u_j, \u_j'), 1 \le j < K$, of distinct vectors in $\cal
U$ such that the differences $\u_j - \u_j'$ are linearly
independent and $\w^*(\u_j - \u_j') = 0, 1 \le j < K$. For $\w$
not proportional to $\w^*$, at least one (but not necessary all)
of the above equalities will break. Therefore all the $\w\u, \u
\in \cal U$ may be not distinct, even if $\w$ is restricted to
${\cal W} \setminus {\cal W}'$. But the set of $\w$ for which this
property is not true anymore is the union of a finite number of
linear subspaces of dimension $K-1$ of $\mathbb{R}^K$ and thus is
not dense in $\mathbb{R}^K$. Therefore for most of the $\w \in
{\cal W} \setminus {\cal W}'$, the $\w\u, \u \in \cal U$ are all
distinct.

The pdf of $\w\S$ can be written as
\begin{equation}
\label{e:density'} p(y) = \sum_{\u \in \cal U} P(\U = \u)
\frac{1}{\sigma} \Phi\Big(\frac{y-\w\u}{\sigma}\Big)\enspace;
\end{equation}
but some of the $\w\u, \u \in \cal U $ can be arbitrarily close to
each other. In this case it is of interest to group the
corresponding terms in~(\ref{e:density'}) together. Thus we
rewrite $p(y)$ as
\[
p(y) = \sum_{n=1}^N \sum_{\u \in {\cal V}_n} P(\u) \Big[\sum_{\u
\in {\cal V}_n} \frac{P(\u)}{\sum_{\u \in {\cal V}_n} P(\u)}
\frac{1}{\sigma} \Phi\Big(\frac{y-\w\u}{\sigma}\Big)\Big]\enspace,
\]
where ${\cal V}_1, \dots, {\cal V}_N$ is a partition of $\cal U$. This
pdf is still of the form~(\ref{e:density}) with
\[
\pi_n = \sum_{\u \in {\cal V}_n} P(\u), \quad K_n(y) = \sum_{\u
\in {\cal V}_n} \frac{P(\u)}{\pi_n} \frac{1}{\sigma}
\Phi\Big(\frac{y-\w\u}{\sigma}\Big) .
\]
The partition ${\cal V}_1, \dots, {\cal V}_N$ can and should be
chosen so that
\[
d(\w) \triangleq \min_{1 \le n \ne m \le N} \min_{\u \in {\cal
V}_n, \u' \in {\cal V}_m} |\w\u - \w\u'|\enspace,
\]
is bounded below by some given positive number. To this end, note
that, as is shown in the proof of Lemma~\ref{lemma2}, $\w^*$ is
associated with a partition ${\cal U}_0^*, \dots, {\cal U}_r^*$ of
$\cal U$ such that $\w^*\u$ take the same value for all $\u \in
{\cal U}_j^*$ ($1 \le j \le r^*$), and the values associated with
different ${\cal U}_j^*$ and the $\w^*\u, \u \in {\cal U}_0^*$,
are all distinct. Thus $\inf_{\w\in\cal W} |\w\u - \w\u'| \ge
\delta$ for some $\delta > 0$ for all $\u \ne \u'$ and $\u, \u'$
do not belong to a same ${\cal U}_j^*, j = 1, \dots, r^*$.
Therefore, the partition $\{{\cal V}_1, \dots, {\cal V}_N\} =
\{\{\u\}, \u \in {\cal U}_0^*, {\cal U}_1^*, \dots, {\cal
U}_r^*\}$ satisfies $d(\w) \ge \delta, \forall \w \in \cal W$. We
then refine this partition by splitting one of the sets ${\cal
U}_j^*, j = 1, \dots, r^*$ into two subsets. The splitting rule is
as follows: for each ${\cal U}_j^*$ arrange the $\w\u, \u \in
{\cal U}_j^*$ in ascending order and look for the maximum gap
between two consecutive values. The set ${\cal U}_j^*$ that
produces the largest gap will be split and the splitting is done
at the gap. For $\w \in {\cal W} \setminus {\cal W}'$, this
maximum gap can be bounded below by a positive number $\delta'$
(noting that there is only a finite number of elements in each
${\cal U}_j^*$); hence for the refined partition, $d(\w) \ge
\min(\delta,\delta')$. Of course, the partition constructed this
way depends on $\w$, but there can be only a finite number of
possible partitions. Hence, one can find a finite number of
subsets ${\cal W}_1, \dots, {\cal W}_q$ which cover ${\cal W}
\setminus {\cal W}'$, each of which is associated with a partition
of $\cal U$ such that the corresponding $d(\w)$ is bounded below
by $\min(\delta,\delta')$ for all $\w$ in this subset. In the
following we shall restrict $\w$ to one such subset, ${\cal W}_p$
say, and we denote by ${\cal V}_1, \dots, {\cal V}_N$ the
associated partition.

We now apply the Lemma~\ref{lemma1} with $\pi_n, K_n, n = 1, \dots, N$
defined as above and with the sets $\Omega_n$ defined by
\[
\Omega_n \triangleq \{y : \min_{\u \in {\cal V}_n}|y - \w\u| <
d(\w)/2\}.
\]
Then we have, writing $d$ in place of $d(\w)$ for short,
\[
\epsilon_n \le 1 - \int_{-d/(2\sigma)}^{-d/(2\sigma)} \Phi(x) dx =
\erfc\Big(\frac{d}{2\sqrt2\sigma}\Big)
\]
\[
\epsilon_n' = \sum_{\u \in {\cal V}_n} \frac{P(\u)}{\pi_n}
\int_{\mathbb{R}\setminus\Omega_n} \frac{1}{\sigma}
\Phi\Big(\frac{y-\w\u}{\sigma}\Big) \log \frac{\sup K_n}{K_n(y)}
dy.
\]
In each term in the sum in that last right hand side, one applies the
bound
\[
\frac{\sup K_n}{K_n(y)} \le \frac{\sigma\sup
K_n}{[P(\u)/\pi_n]\Phi[(y-\w\u)/\sigma]}
\]
which yields,
\[
\displaylines{
\epsilon_n' \le \sum_{\u \in {\cal V}_n} \frac{P(\u)}{\pi_n}
\int_{[x|>d/(2\sigma)} \Phi(x) \log \frac{\sigma\sup
K_n}{[P(\u)/\pi_n]\Phi(x)} dx
\hfill\cr\hfill
{}= \Big[\log\sup(\sigma K_n) -
\sum_{\u\in{\cal V}_n} \frac{P(\u)}{\pi_n} \log\frac{P(\u)}{\pi_n}
\Big]\erfc\Big(\frac{d}{2\sqrt2\sigma}\Big) \hfill\cr\hfill {}+
H(\Phi) - H_{d/\sigma}(\Phi) . }
\]
Therefore, putting $h_n = -\sum_{\u\in{\cal V}_n} [P(\u)/\pi_n]
\log[P(\u)/\pi_n]$ and noting that $\sup(\sigma K_n) \le \sup\Phi
= (2\pi)^{-1/2}$, one gets
\[
\displaylines{ \sum_{n=1}^N \pi_n \epsilon_n' + \sum_{n=1}^N \pi_n
\Big[ \log\Big(\frac{\max_{1 \le m \le N} \sup K_m}{\pi_n \sup
K_n}\Big) + 1\Big] \epsilon_n \le \hfill\cr\hfill \Big[1 -
\frac{\log(2\pi)}{2} + \sum_{n=1}^N \pi_n h_n\Big]
\erfc\Big(\frac{d}{2\sqrt2\sigma}\Big) + H(\Phi) -
H_{d/\sigma}(\Phi) }
\]\enspace.
Since $d = d(\w) \ge \min(\delta,\delta'), \forall \w \in {\cal
W}_p$, the last inequality shows that for any $\eta > 0$,
\[
H(p) \ge \sum_{n=1}^N \pi_n H(K_n) + h(\pi) - \eta,
\quad \forall \w \in {\cal W}_p,
\]
for $\sigma$ small enough. On the other hand, since $\log x \le x -
1$,
\[
\int \frac{1}{\sigma} \Phi\Big(\frac{y-\w\u}{\sigma}\Big) \log
\frac{K_n(y)}{\Phi[(y-\w\u)/\sigma]/\sigma} dy \le 0.
\]
Multiplying both members of the above inequality by $P(\u)/\pi_n$
and summing up with respect to $\u \in {\cal V}_n$, one gets
$H(\Phi) + \log\sigma - H(K_n) \le 0$. Therefore
\[
H(p) \ge H(\Phi) + \log\sigma + h(\pi) - \eta\enspace.
\]
But by construction $h(\pi) > h(\w^*\U)$ (see the proof of
Lemma~\ref{lemma2}); therefore, taking $\eta < h(\pi) -
h(\w^*\U)$, one sees that for $\sigma$ small enough $H(\w\S) =
H(p) > H(\w^*\S)$ for all $\w \in {\cal W}_p$. Since this is true
for all $p = 1, \dots, q$, we conclude as before that $H(\w\S)$
admits a local minimum in $\cal W'$.

\qed

\begin{biography}
[{\includegraphics[width=1in,height=1.25in,clip,keepaspectratio]{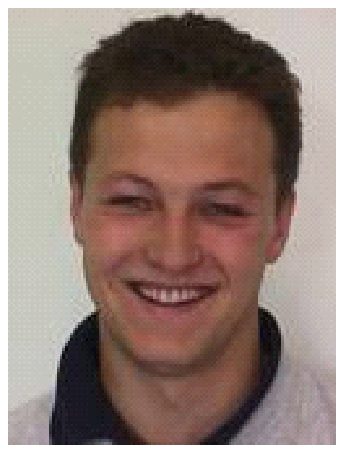}}]{Fr\'ed\'eric
Vrins} was born in Uccle, Belgium, in 1979. He received the MS
degree in mechatronics engineering and the DEA degree in Applied
Sciences from the Universit\'e catholique de Louvain (Belgium) in
2002 and 2004, respectively. He is currently working towards the
PhD degree in the UCL Machine Learning Group. His research
interests are blind source separation, independent component
analysis, Shannon and Renyi entropies, mutual information and
information theory in adaptive signal
processing.
He is member of the program
committee of ICA 2006.
\end{biography}
\begin{biography}[{\includegraphics[width=1in,height=1.25in,clip,keepaspectratio]{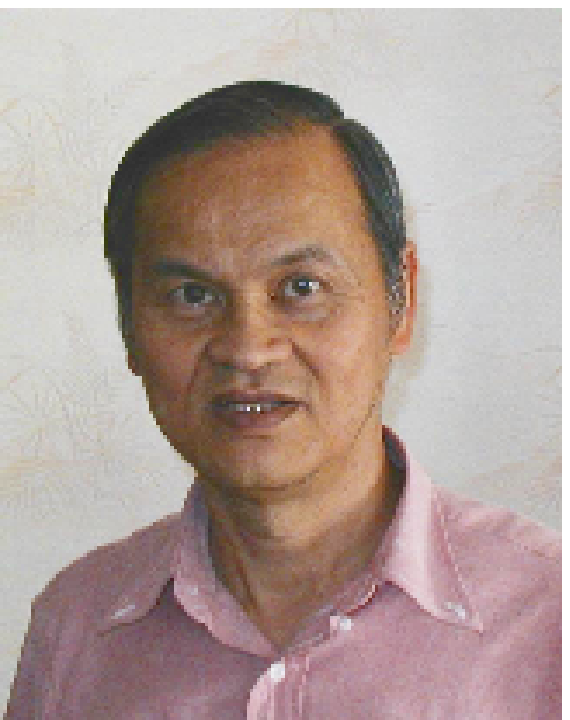}}]{Dinh-Tuan Pham}
 was born in Hanoï, VietNam, on February 10, 1945. He is
graduated from the Engineering School of Applied Mathematics and
Computer Science (ENSIMAG) of the Polytechnic Institute of
Grenoble in 1968. He received the Ph. D. degree in Statistics in
1975 from the University of Grenoble. He was a Postdoctoral Fellow
at Berkeley (Department of Statistics) in 1977-1978 and a Visiting
Professor at Indiana University (Department of Mathematics) at
Bloomington in 1979-1980. He is currently Director of Research at
the French Centre National de Recherche Scientifique (C.N.R.S).
His researches include time series analysis, signal modelling,
blind source separation, nonlinear (particle) filtering and
biomedical signal processing.

\end{biography}
\begin{biography}
[{\includegraphics[width=1in,height=1.25in,clip,keepaspectratio]{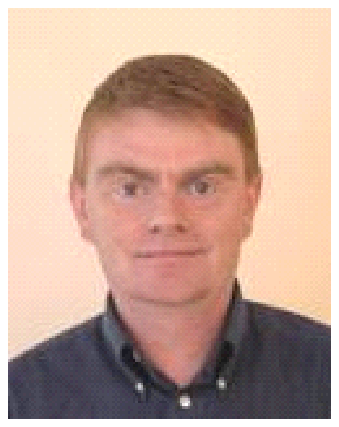}}]{Michel
Verleysen} was born in 1965 in Belgium. He received the M.S. and
Ph.D. degrees in electrical engineering from the Universit\'e
catholique de Louvain (Belgium) in 1987 and 1992, respectively. He
was an Invited Professor at the Swiss E.P.F.L. (Ecole
Polytechnique F\'ed\'erale de Lausanne, Switzerland) in 1992, at
the Universit\'e d'Evry Val d'Essonne (France) in 2001, and at the
Universit\'e Paris I-Panth\'eon-Sorbonne in 2002, 2003 and 2004.
He is now Research Director of the Belgian F.N.R.S. (Fonds
National de la Recherche Scientique) and Lecturer at the
Universit\'e catholique de Louvain. He is editor-in-chief of the
Neural Processing Letters journal and chairman of the annual ESANN
conference (European Symposium on Artificial Neural Networks); he
is associate editor of the IEEE Trans. Neural Networks journal,
and member of the editorial board and program committee of several
journals and conferences on neural networks and learning. He is
author or co-author of about 200 scientific papers in
international journals and books or communications to conferences
with reviewing committee. He is the co-author of the scientific
popularization book on artificial neural networks in the series
"Que Sais-Je?", in French. His research interests artificial
neural networks, self-organization, time-series forecasting,
nonlinear statistics, adaptive signal processing,
information-theoretic learning and biomedical data and signal
analysis.
\end{biography}
\vfill
\end{document}